\documentclass{ifacconf}

\usepackage{graphicx}      
\usepackage{natbib} 
\usepackage{amsmath}
\usepackage{amssymb}     
\usepackage{subfigure}   
\usepackage[latin1]{inputenc}
\usepackage[T1]{fontenc}
\newtheorem{remark}{Remark}  
       
\begin{document}
\begin{frontmatter}

\title{On short-term traffic flow forecasting \\ and its reliability
} 


\author[First]{Hassane Aboua\"{\i}ssa} 
\author[Second,Fourth]{Michel Fliess} 
\author[Third,Fourth,Fifth]{C\'{e}dric Join}

\address[First]{Laboratoire de G\'{e}nie Informatique et d'Automatique de l'Artois \\ (LGI2A, EA 3926), Universit\'{e} d'Artois, 62400 B\'{e}thune, France \\ (e-mail: hassane.abouaissa@univ-artois.fr)}
\address[Second]{LIX (CNRS, UMR 7161), \'Ecole polytechnique, 91128 Palaiseau, France (e-mail$:$ Michel.Fliess@polytechnique.edu)}
\address[Third]{CRAN (CNRS, UMR 7039), Universit\'{e} de Lorraine, BP 239, 54506 Vand{\oe}uvre-l\`{e}s-Nancy, France \\ (e-mail: cedric.join@univ-lorraine.fr)}
\address[Fourth]{AL.I.E.N. (ALg\`{e}bre pour Identification \& Estimation Num\'{e}riques), 24-30 rue Lionnois, BP 60120, 54003 Nancy, France \\ (e-mail: \{michel.fliess, cedric.join\}@alien-sas.com)}
\address[Fifth]{Projet Non-A, INRIA Lille -- Nord-Europe, France}

\begin{abstract}             
Recent advances in time series, where deterministic and stochastic modelings as well as the storage and analysis of big data are useless,  permit a new approach to short-term traffic flow forecasting and to its reliability, \textit{i.e.}, to the traffic volatility. Several convincing 
computer simulations, which utilize concrete data, are presented and discussed. 
\end{abstract}

\end{frontmatter}

\section{Introduction}
We recently proposed a new feedback control law for ramp metering (\cite{sofia}), which is based on the most fruitful \emph{model-free control} setting (\cite{ijc}). It has not only been patented but also
successfully tested in 2015 on a highway in northern France.\footnote{See, \textit{e.g.}, the newspaper \textit{La Voix du Nord}, 2 December 2015, p. 3.} It will soon be implemented on a larger scale. We are therefore lead to study another important topic for intelligent transportation systems, \textit{i.e.}, short-term traffic flow forecasting:
it plays a key r\^{o}le in the planning and development of traffic management. This importance explains the extensive literature on this subject since at least thirty years. Several surveys (see, \textit{e.g.}, \cite{bol,chang,lippi,smith,vl}) 
provide useful informations on the various approaches which have been already employed: regression analysis, time series, expert systems, artificial neural networks, fuzzy logic, etc. We follow here another road, \textit{i.e.}, a new approach to time series (\cite{perp,esaim,ipag,douai,agadir}):
\begin{itemize}
\item A quite recent theorem due to \cite{cartier} yields the most important notions of \emph{trends} and \emph{quick fluctuations}, which do not seem to have any analogue in other theoretical approaches. Among those existing
approaches, the dominant one today has been developed for econometric goals (see, \textit{e.g.}, \cite{melard}, \cite{tsay}, and \cite{meuriot} for some historical and epistemological issues). It is quite popular in traffic flow forecasting.
\item Although its origin lies in financial engineering, it has been recently applied for short-term meteorological forecasts for the purpose of renewable energy management (\cite{paris,voyant,troyes}).
\item Like in model-free control (\cite{ijc}), no deterministic or probabilistic mathematical modeling is needed. Moreover the storage and analysis of \emph{big data} is useless.Those facts open new perspectives to intelligent knowledge-based systems.
\end{itemize} 
The reliability of those computations should nevertheless be examined, at least for a better risk understanding. 
This subject, which is crucial for any type 
of approach, has been much less studied (see, \textit{e.g.}, \cite{guo,laflamme,zhang}, and the references therein). This risk may of course be studied via the concept of \emph{volatility}, which may be found everywhere in finance 
(see, \textit{e.g.}, \cite{tsay,wilmott}). The strong attacks against the very concept of volatility seem to have been ignored in the community studying intelligent transportation systems. We are thus reproducing the following quote from 
\cite{douai}.
\cite{wilmott} (chap. 49, p. 813) writes: \textit{Quite frankly, we do not know what volatility currently is,
never mind what it may be in the future}. 
The lack moreover of any precise mathematical definition leads to
multiple ways for computing volatility which are by no means
equivalent and might even be sometimes misleading (see,
\textit{e.g.}, \cite{gol}). Our theoretical formalism and the
corresponding computer simulations will confirm what most
practitioners already know. It is well
expressed by \cite{gunn} (p. 49): 
\textit{Volatility is not only referring to something that
fluctuates sharply up and down but is also referring to something
that moves sharply in a sustained direction}. Let us stress that in econometrics and in financial engineering the notion of volatility is usually examined via the \emph{returns} of financial assets. This setting seems to be pointless in the context 
of traffic flow. Defining the volatility directly from the time 
series (see also \cite{agadir}) makes much more sense.

Our viewpoint on time series is sketched in Section \ref{basics}. Section \ref{persis} investigates the fundamental notion of \emph{persistence}. The forecasting techniques for the traffic flow on a 
French highway and the corresponding computer experiments are discussed in Section \ref{cast}. Short concluding remarks may be found in Section \ref{conclusion}.

\section{Revisiting time series}\label{basics}
\subsection{Time series via nonstandard analysis}\label{cartierperrin}
Take the time
interval $[0, 1] \subset \mathbb{R}$ and introduce as often in
\emph{nonstandard analysis} (see, \textit{e.g.}, (\cite{lobry,perp,esaim}), and some of the references therein, for basics in nonstandard analysis) for the infinitesimal sampling
$${\mathfrak{T}} = \{ 0 = t_0 < t_1 < \dots < t_\nu = 1 \}$$
where $t_{i+1} - t_{i}$, $0 \leq i < \nu$, is {\em infinitesimal},
{\it i.e.}, ``very small''. A
\emph{time series} $X(t)$ is a function $X: {\mathfrak{T}}
\rightarrow \mathbb{R}$.

A time series ${\mathcal{X}}: {\mathfrak{T}} \rightarrow \mathbb{R}$
is said to be {\em quickly fluctuating}, or {\em oscillating}, if,
and only if, the integral $\int_A {\mathcal{X}} dm$ is
infinitesimal, \textit{i.e.}, very small, for any \emph{appreciable} interval, \textit{i.e.}, an interval which is neither very small nor very large.

According to a theorem due to 
\cite{cartier} the following additive decomposition holds for any time series $X$, which satisfies a weak integrability condition,
\begin{equation}\label{decomposition}
\boxed{X(t) = E(X)(t) + X_{\tiny{\rm fluctuation}}(t)}
\end{equation}
where
\begin{itemize}
\item the \emph{mean}, or \emph{average}, $E(X)(t)$ is ``quite smooth.'',
\item $X_{\tiny{\rm fluctuation}}(t)$ is quickly fluctuating.
\end{itemize}
The decomposition \eqref{decomposition} is unique up to an
infinitesimal.

\subsection{On the numerical differentiation of a noisy signal}\label{ins} 
Let us start with the first degree polynomial time function $p_{1} (\tau)
= a_0 + a_1 \tau$, $\tau \geq 0$, $a_0, a_1 \in \mathbb{R}$. Rewrite
thanks to classic operational calculus with respect to the variable $\tau$ (see, \textit{e.g.},
\cite{yosida}) $p_1$ as $P_1 = \frac{a_0}{s} +
\frac{a_1}{s^2}$. Multiply both sides by $s^2$:
\begin{equation}\label{1}
s^2 P_1 = a_0 s + a_1
\end{equation}
Take the derivative of both sides with respect to $s$, which
corresponds in the time domain to the multiplication by $- t$:
\begin{equation}\label{2}
s^2 \frac{d P_1}{ds} + 2s P_1 = a_0
\end{equation}
The coefficients $a_0, a_1$ are obtained via the triangular system
of equations (\ref{1})-(\ref{2}). We get rid of the time
derivatives, i.e., of $s P_1$, $s^2 P_1$, and $s^2 \frac{d
P_1}{ds}$, by multiplying both sides of Equations
(\ref{1})-(\ref{2}) by $s^{ - n}$, $n \geq 2$. The corresponding
iterated time integrals are low pass filters which attenuate the
corrupting noises (see \cite{bruit} for an explanation). A quite short time window is sufficient for
obtaining accurate values of $a_0$, $a_1$. Note that estimating $a_0$ yields the trend.

The extension to polynomial functions of higher degree is
straightforward. For derivative estimates up to some finite order
of a given smooth function $f: [0, + \infty) \to \mathbb{R}$, take a
suitable truncated Taylor expansion around a given time instant
$t_0$, and apply the previous computations. Resetting  and utilizing
sliding time windows permit to estimate derivatives of various
orders at any sampled time instant.
\begin{remark}
See (\cite{easy,NumDiff,sira}) for more details.
\end{remark}
\subsection{Forecasting}
Set the following forecast $X_{\text{est}}(t + \Delta T)$, where $\Delta T > 0$ is not too ``large'',
\begin{equation}\label{delta}
X_{\text{forecast}}(t + \Delta T) = E(X)(t) + \left[\frac{d E(X)(t)}{dt}\right]_e \Delta T
\end{equation}
where $E(X)(t)$ and $\left[\frac{d E(X)(t)}{dt}\right]_e$ are estimated like $a_0$ and $a_1$ in Section \ref{ins}. Let us stress that what we predict is the trend and not the quick fluctuations (see also \cite{perp,agadir,paris,voyant}).
\subsection{Volatility}\label{volat}
Contrarily to our previous approach via returns (\cite{douai,agadir}), we use here the difference $X(t) - E(X)(t)$ between the time series and its trend. 
If this difference is square integrable, \textit{i.e.}, if $(X(t) - E(X)(t))^2$ is integrable, \emph{volatility} is defined via the following standard deviation type formula:
\begin{eqnarray*}
{{\text{\bf vol}}} (X)(t)  &=& \sqrt{E\left(X - E(X)\right)^2} \\ 
&\simeq& \sqrt{E (X^2) - E(X)^2}
\end{eqnarray*}

\section{Persistence} \label{persis}
\subsection{Definition}
The \emph{persistence} method is the simplest way of producing a forecast. It assumes that the conditions at the time of the forecast will not change, \textit{i.e.},
\begin{equation}\label{pers}
X_{\text{forecast}}(t + \Delta T) = X(t)
\end{equation}
\subsection{Scaled Persistence}
\emph{Scaled persistence}, which is often encountered in meteorology (see, \textit{e.g.}, (\cite{lauret}), and (\cite{paris,voyant}))  improves Formula \eqref{pers} by writing
\begin{equation}\label{scapers}
X_{\text{Pe}}(t + \Delta T) = E(X)(t) \times S_c(t)
\end{equation}
where 
\begin{itemize}
\item $E(X)(t)$ is estimated like $a_0$ in Section \ref{cartierperrin},
\item the scaling factor $S_c(t)$ will be made precise according to the situation,
\item contrarily to \eqref{pers} quick fluctuations are disregarded and the trend is emphasized.
\end{itemize}

\section{Case study}\label{cast}
\subsection{Description }
\begin{figure*}[htbp]
\begin{center}
\includegraphics[width=13.5cm]{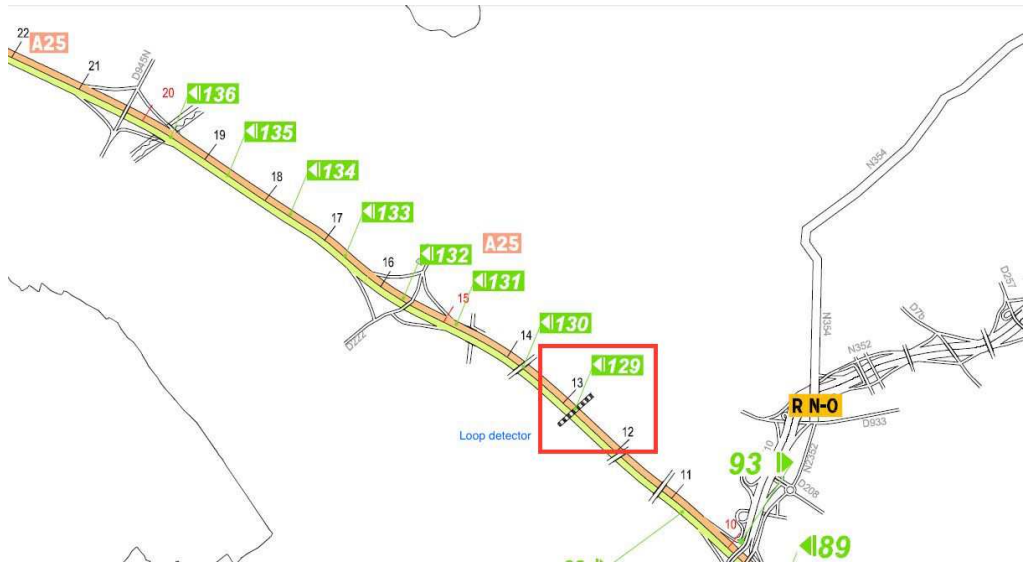}
\caption{Our highway section}
\label{autoroute}
\end{center}
\end{figure*}

\begin{figure*}
\begin{center}
\subfigure[From 1 to 30 June 2014]{
\resizebox*{5.915cm}{!}{\includegraphics{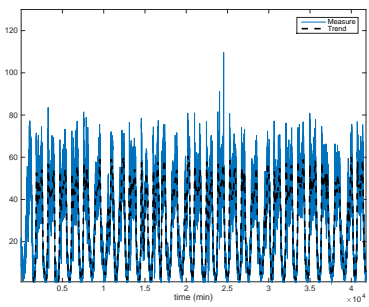}}}
\subfigure[Zoom 1]{
\resizebox*{5.915cm}{!}{\includegraphics{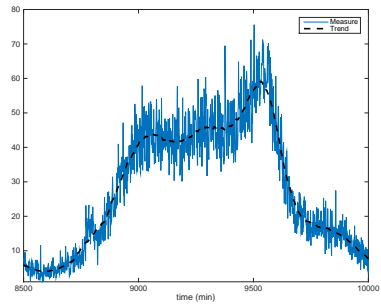}}}%
\subfigure[Zoom 2]{
\resizebox*{5.915cm}{!}{\includegraphics{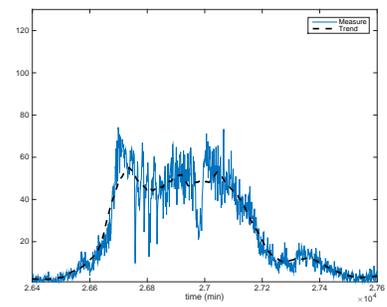}}}%
\caption{Measures and trend}%
\label{T}
\end{center}
\end{figure*}

Consider a section of the highway A25 from Dunkirk (\textit{Dunkerque} in French) to Lille (see Figure \ref{autoroute}). There are two lanes on this section, and about 
$900$m between two measurements stations. Congestions often occur.
The traffic volume, the occupation rate and and the mean vehicle speed, which yield excellent traffic characterizations, are measured. We focus here on the traffic volume $Q(t)$, in veh/min. 
It is registered every minute from 1 to 30 June 2014, and
displayed in Figure \ref{T}-(a). Two single days are detailed in Figures  \ref{T}-(b) and \ref{T}-(c).
In all those Figures the trend is also drawn. It is computed by using 100 points and the following non-causal moving average 
{\tiny \begin{equation}\label{mean}
\text{mean}(Q(t-49),...,Q(t+50)) = \frac{Q(t-49) + \dots + Q(t+50)}{100} 
\end{equation}}

\subsection{Forecastings}
Let us emphasize that forecasting errors will be defined with respect to the trend derived from \eqref{mean}. Three forecast horizons are considered$:$ $5$, $15$, and $60$ minutes. Set $X(t) = Q(t)$. 
The term $E(Q)(t)$ in \eqref{delta} and \eqref{scapers} are deduced from the causal moving average
{\small
$$
E_{100}(Q)(t) = \text{mean}(Q(t-99),...,Q(t)) = \frac{Q(t-99) + \dots + Q(t)}{100} 
$$}
The scaling factor $S_c(t)$ in \eqref{scapers} is given by 
\begin{equation*}\label{scal}
S_c(t) = \frac{E_{100}(Q)(t - 1\text{day} + \Delta T)}{E_{100}(Q)(t - 1\text{day})}
\end{equation*}
where 
\begin{itemize}
\item $\text{1day} = 60 \times 24 = 1440 \  \text{minutes}$,
\item $\Delta T$ is equal to one of the three following values$:$ 5, 15, 60 minutes. 
\end{itemize}
Then \eqref{delta} and \eqref{scapers} become respectively
\begin{equation}\label{A1}
Q_{\text{A}}(t + \Delta T) = E_{100}(Q)(t) + \left[\frac{d E_{100}(Q)(t)}{dt}\right]_e \Delta T
\end{equation}
and
\begin{equation}\label{scapers100}
Q_{\text{Pe}}(t + \Delta T) = E_{100}(Q)(t) \times \frac{E_{100}(Q)(t - 1\text{day} + \Delta T)}{E_{100}(Q)(t - 1\text{day})}
\end{equation}
Computer experiments show that \eqref{A1} and \eqref{scapers100} suffer respectively from rather large overshoots and undershoots. In order to remedy this annoying fact write \eqref{scapers100} in the form
$$
Q_{\text{Pe}}(t+\Delta t)=E_{100}(Q)(t)+E_{100}(Q)(t)\frac{Sc(t)-1}{\Delta t}\Delta t
$$
It yields the following forecasting equation
\begin{equation}
\label{eqMi}
Q_{\text{MixedForecast}}(t+\Delta t) = E_{100}(Q)(t) +\tilde{a}_1(t) \Delta t
\end{equation}
where $\tilde{a}_1$ is equal to 
\begin{enumerate}
\item $\frac{d E_{100}(Q)(t)}{dt}$ if its module is smaller than the module of
$E_{100}(Q)(t)\frac{Sc(t)-1}{\Delta t}$,
\item $E_{100}(Q)(t)\frac{Sc(t)-1}{\Delta t}$ if not.
\end{enumerate}

\subsection{Computer Experiments} 
\begin{figure*}
\begin{center}
\subfigure[The whole set of data]{
\resizebox*{5.915cm}{!}{\includegraphics{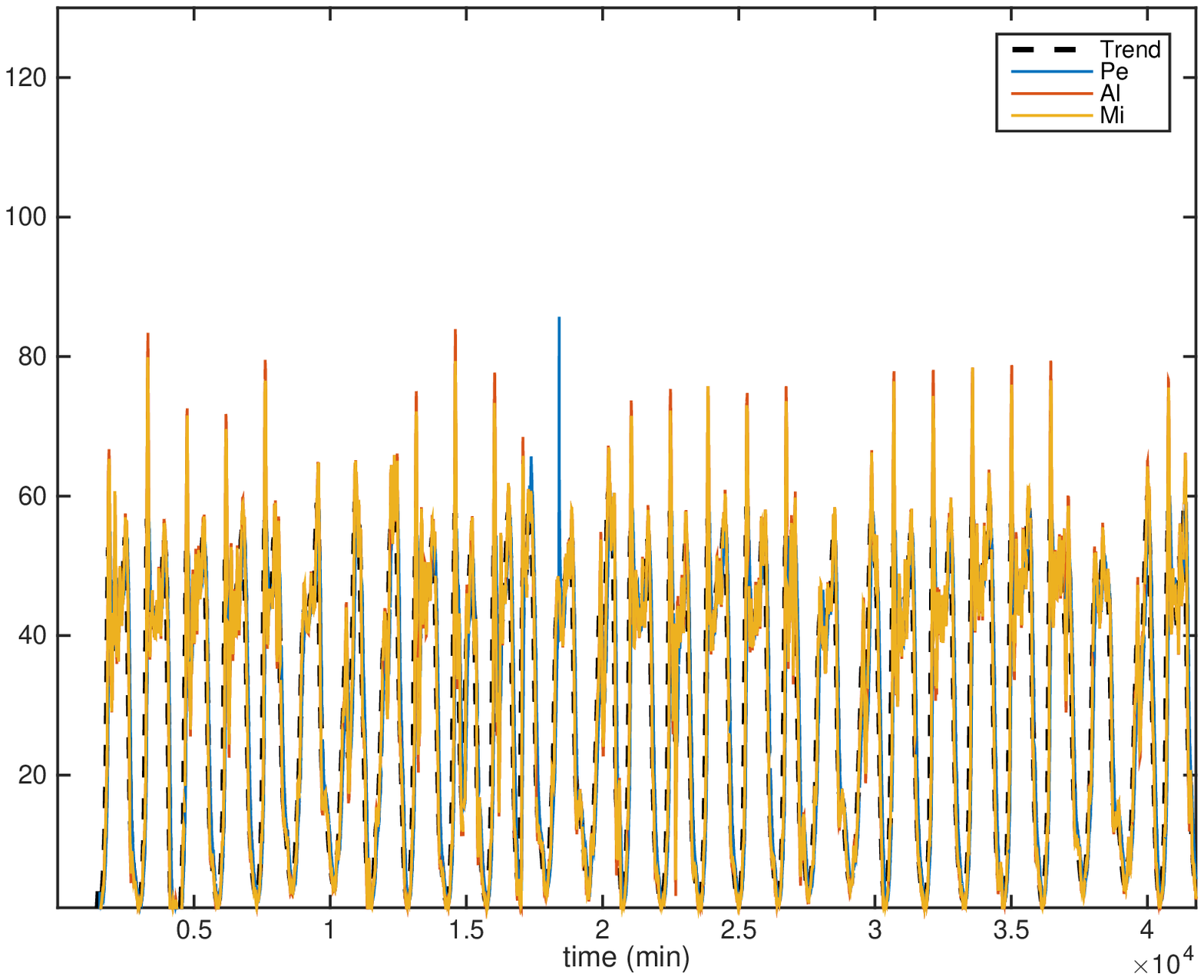}}}%
\subfigure[Zoom 1]{
\resizebox*{5.915cm}{!}{\includegraphics{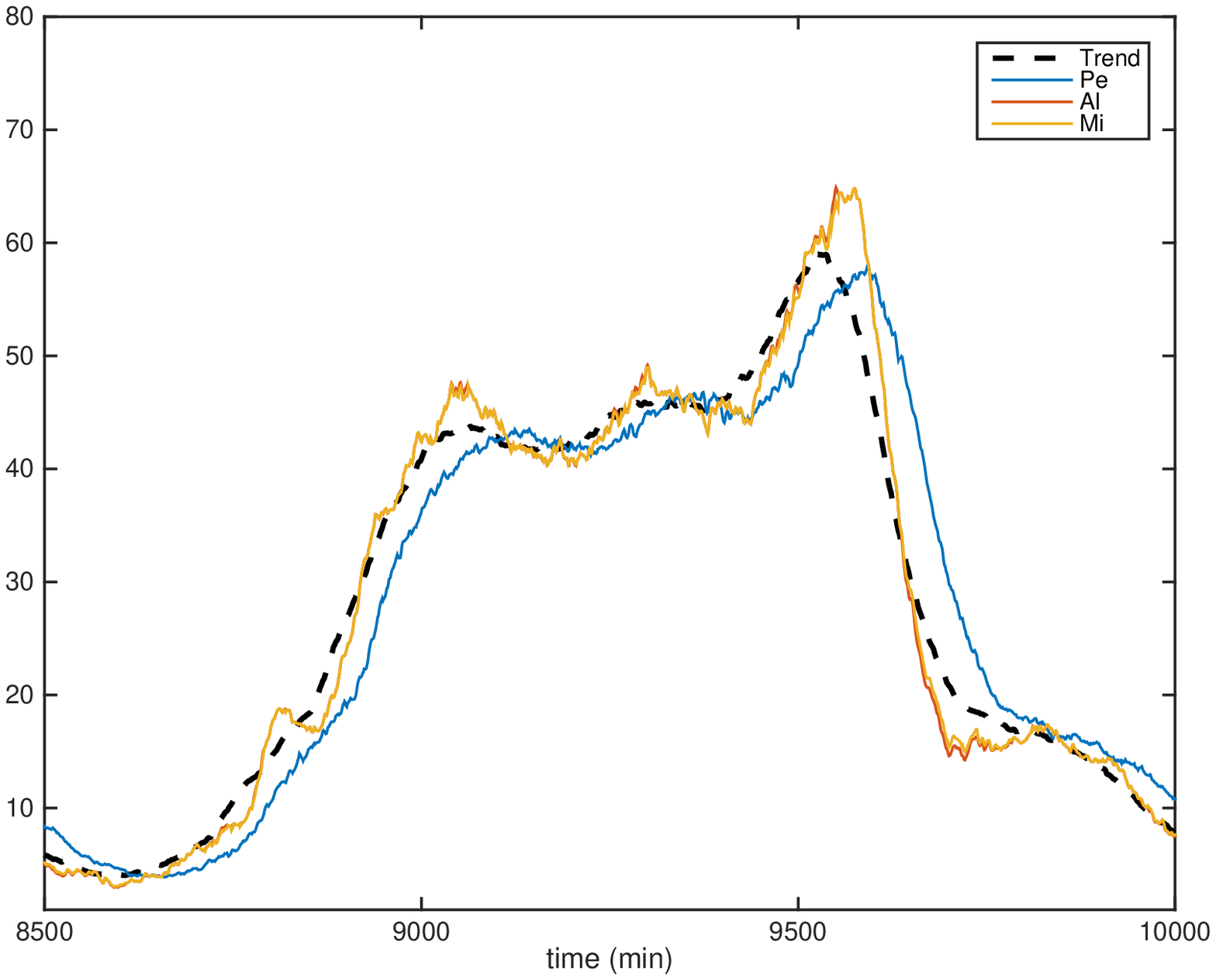}}}%
\subfigure[Zoom 2]{
\resizebox*{5.915cm}{!}{\includegraphics{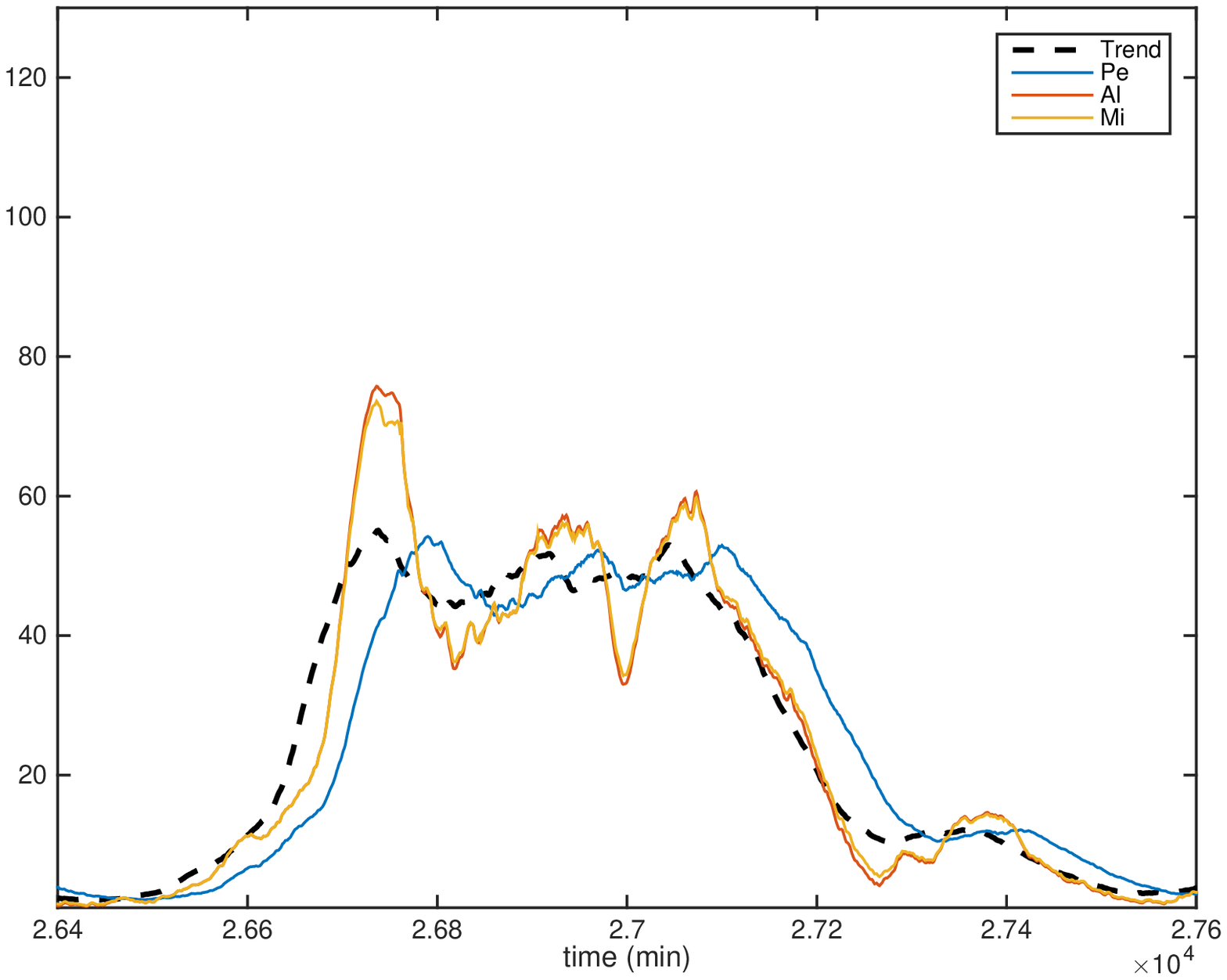}}}%
\caption{5 minutes forecasts}%
\label{P5}
\end{center}
\end{figure*}

\begin{figure*}
\begin{center}
\subfigure[From 1 to 30 June 2014]{
\resizebox*{5.915cm}{!}{\includegraphics{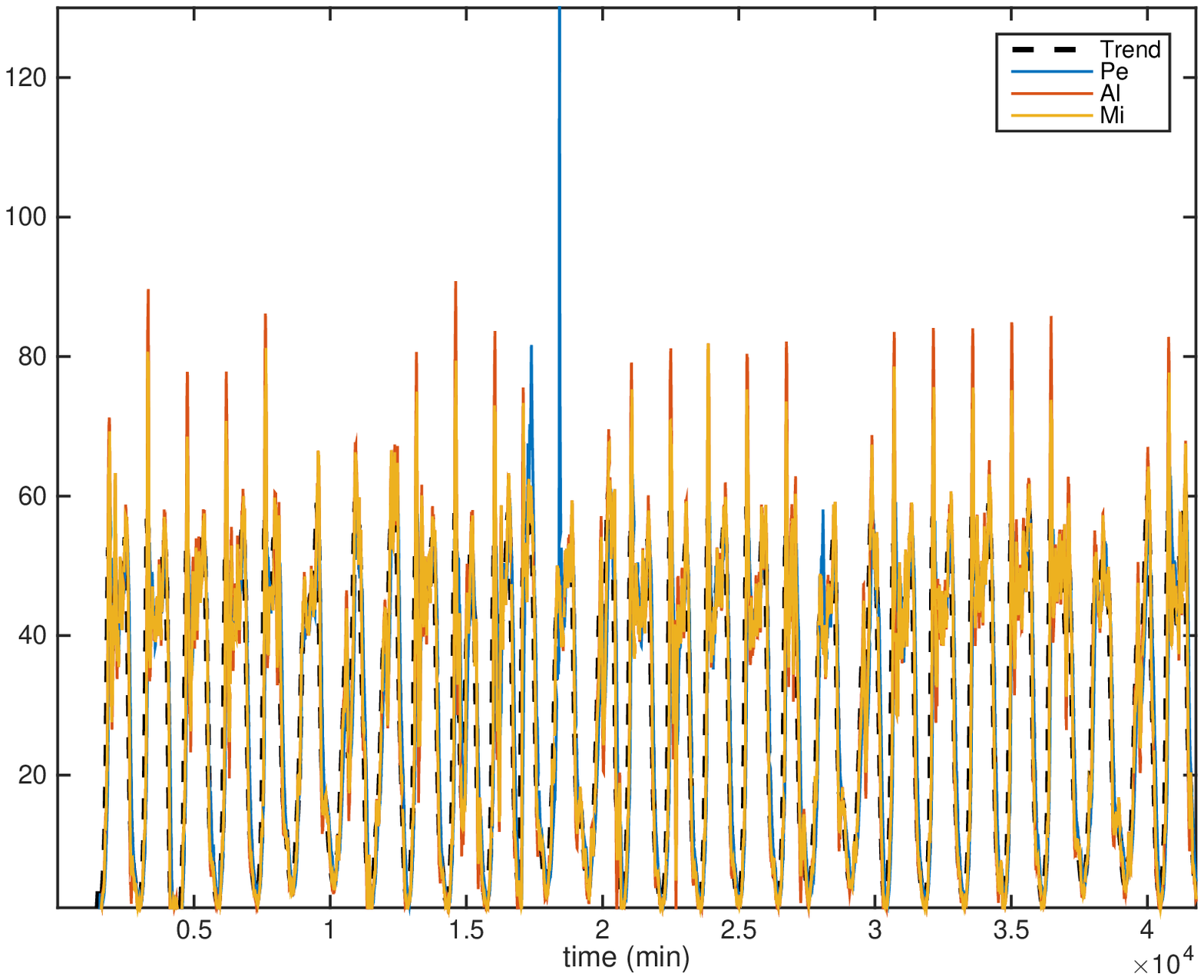}}}
\subfigure[Zoom 1]{
\resizebox*{5.915cm}{!}{\includegraphics{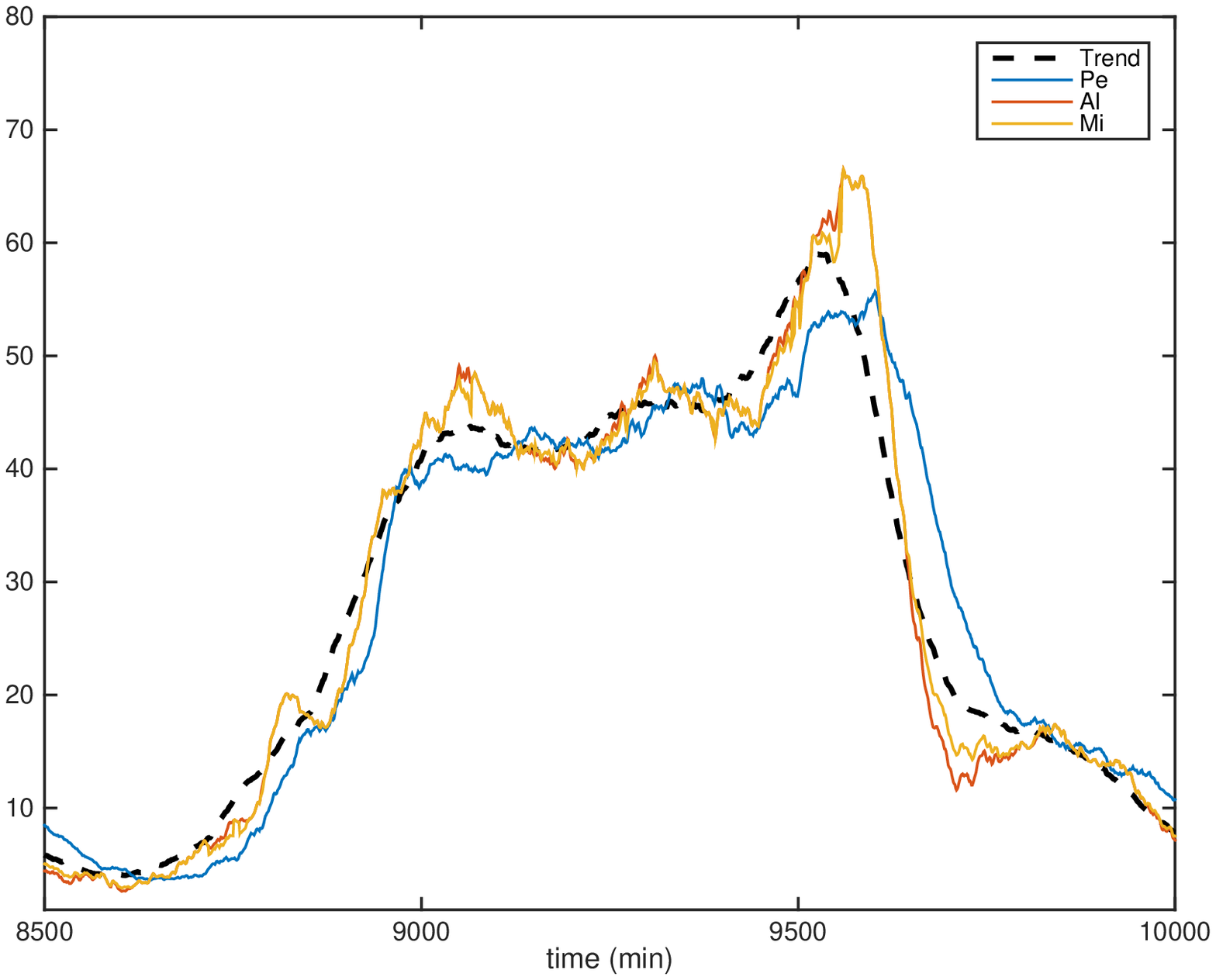}}}%
\subfigure[Zoom 2]{
\resizebox*{5.915cm}{!}{\includegraphics{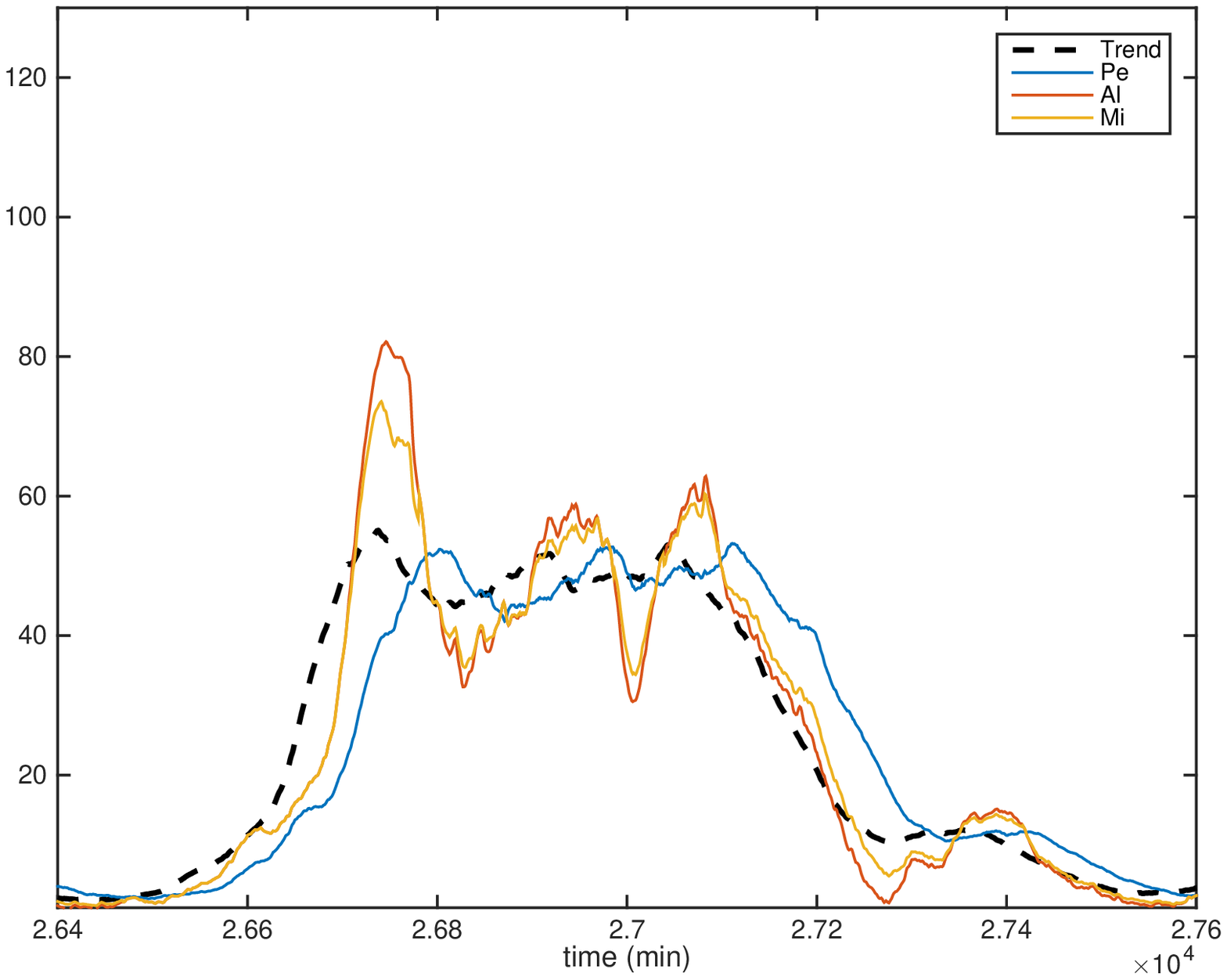}}}%
\caption{15 minutes forecasts}%
\label{P15}
\end{center}
\end{figure*}

\begin{figure*}
\begin{center}
\subfigure[The whole set of data]{
\resizebox*{5.915cm}{!}{\includegraphics{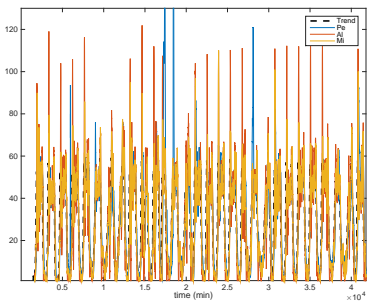}}}
\subfigure[Zoom 1]{
\resizebox*{5.915cm}{!}{\includegraphics{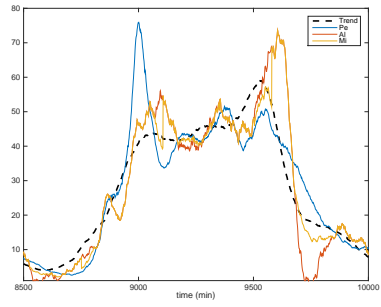}}}%
\subfigure[Zoom 2]{
\resizebox*{5.915cm}{!}{\includegraphics{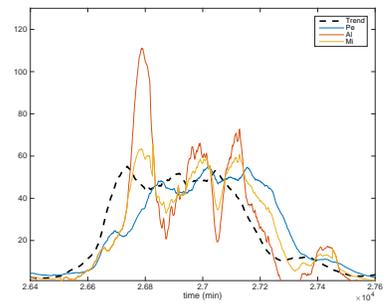}}}%
\caption{60 minutes forecasts}%
\label{P60}
\end{center}
\end{figure*}

Results are displayed in Figures \ref{P5}, \ref{P15} and \ref{P60} . The superiority of the forecasting \eqref{eqMi} is obvious. Table \ref{tb:er}, with its squared errors, provide a quantified comparison of the various approaches.
\begin{table}[hb]
\begin{center}
\caption{$\sum \text{Errors}^2$}\label{tb:er}
\begin{tabular}{cccc}
Horizon & Pe & Al \textit{[gain in \%]} & Mi \textit{[gain in \%]} \\\hline
$t+5\text{min}$ & 2.08e+06 & 1.01e+06 \textit{[105\%]}& 8.75e+05 \textit{[137\%]}\\
$t+15\text{min}$ & 2.64e+06 & 1.7335e+06 \textit{[52\%]}& 1.23e+06 \textit{[114\%]}\\ 
$t+60\text{min}$ &1.15e+07 & 8.47e+06 \textit{[36\%]} & 4.29e+06 \textit{[169\%]}\\ \hline
\end{tabular}
\end{center}
\end{table}

\subsection{Volatility}
\begin{figure*}
\begin{center}
\subfigure[The whole set of data]{
\resizebox*{5.915cm}{!}{\includegraphics{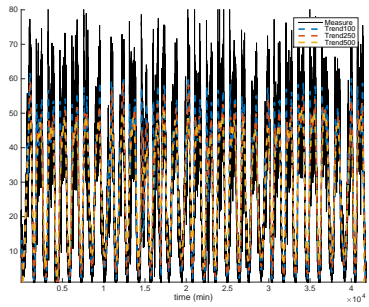}}}
\subfigure[Zoom 1]{
\resizebox*{5.915cm}{!}{\includegraphics{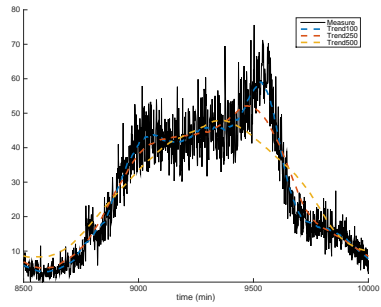}}}%
\subfigure[Zoom 2]{
\resizebox*{5.915cm}{!}{\includegraphics{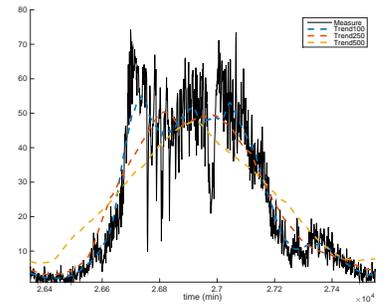}}}%
\caption{Trend and time scales}%
\label{Ts}
\end{center}
\end{figure*}

\begin{figure*}
\begin{center}
\subfigure[The whole set of data]{
\resizebox*{5.915cm}{!}{\includegraphics{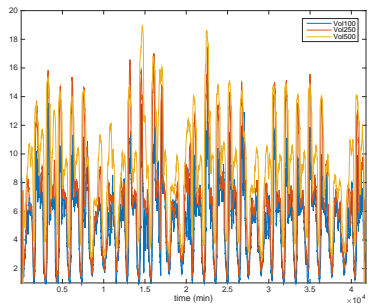}}}
\subfigure[Zoom 1]{
\resizebox*{5.915cm}{!}{\includegraphics{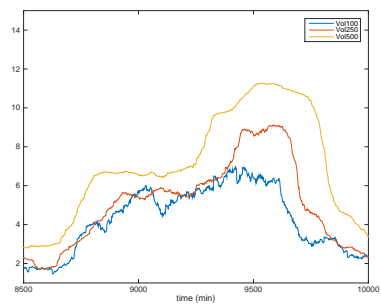}}}%
\subfigure[Zoom 2]{
\resizebox*{5.915cm}{!}{\includegraphics{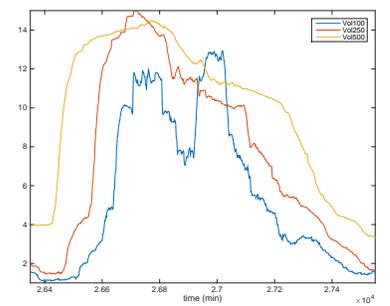}}}%
\caption{Volatility and time scales}%
\label{Vs}
\end{center}
\end{figure*}

\begin{figure*}
\begin{center}
\subfigure[The whole set of data]{
\resizebox*{5.915cm}{!}{\includegraphics{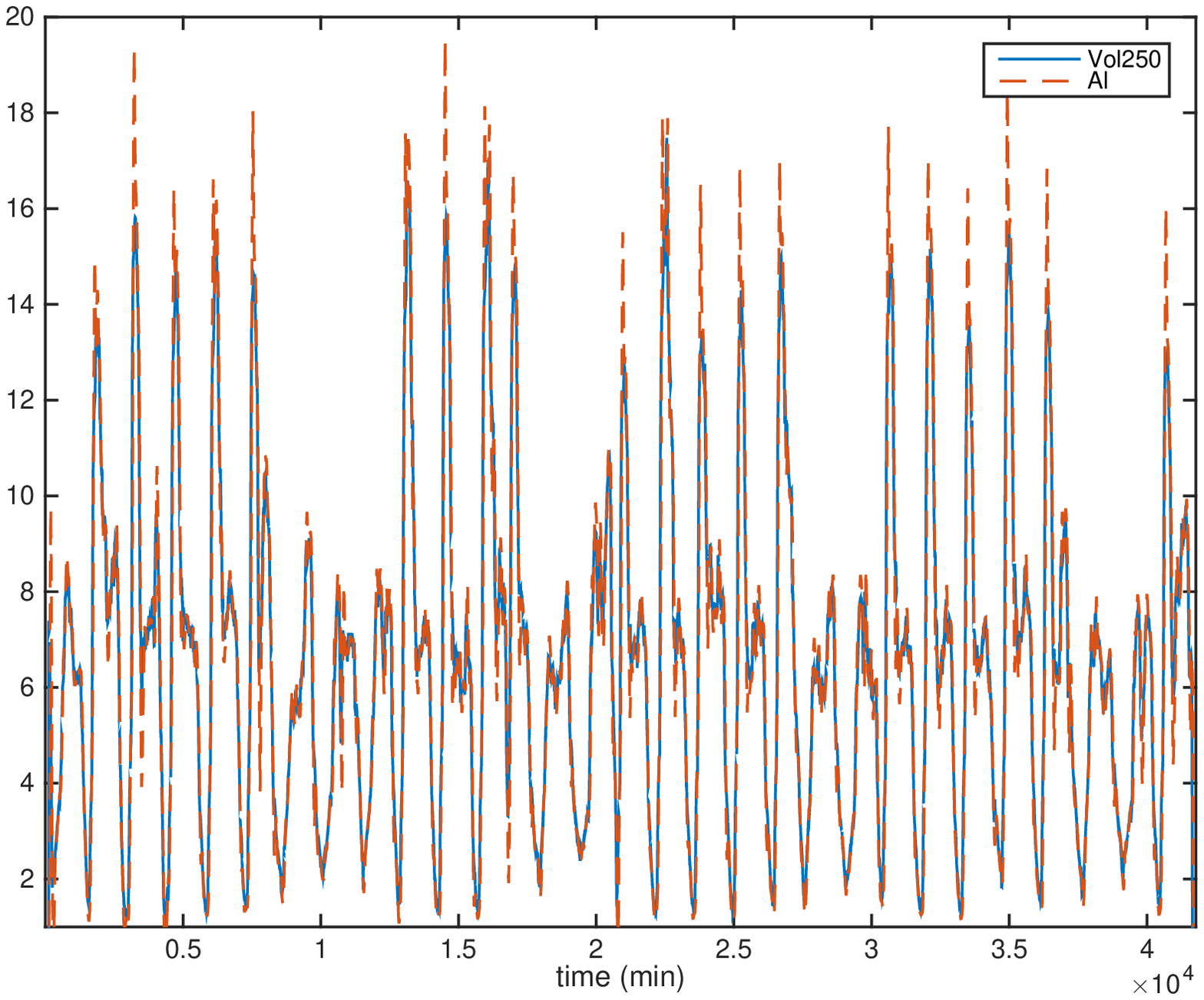}}}
\subfigure[Zoom 1]{
\resizebox*{5.915cm}{!}{\includegraphics{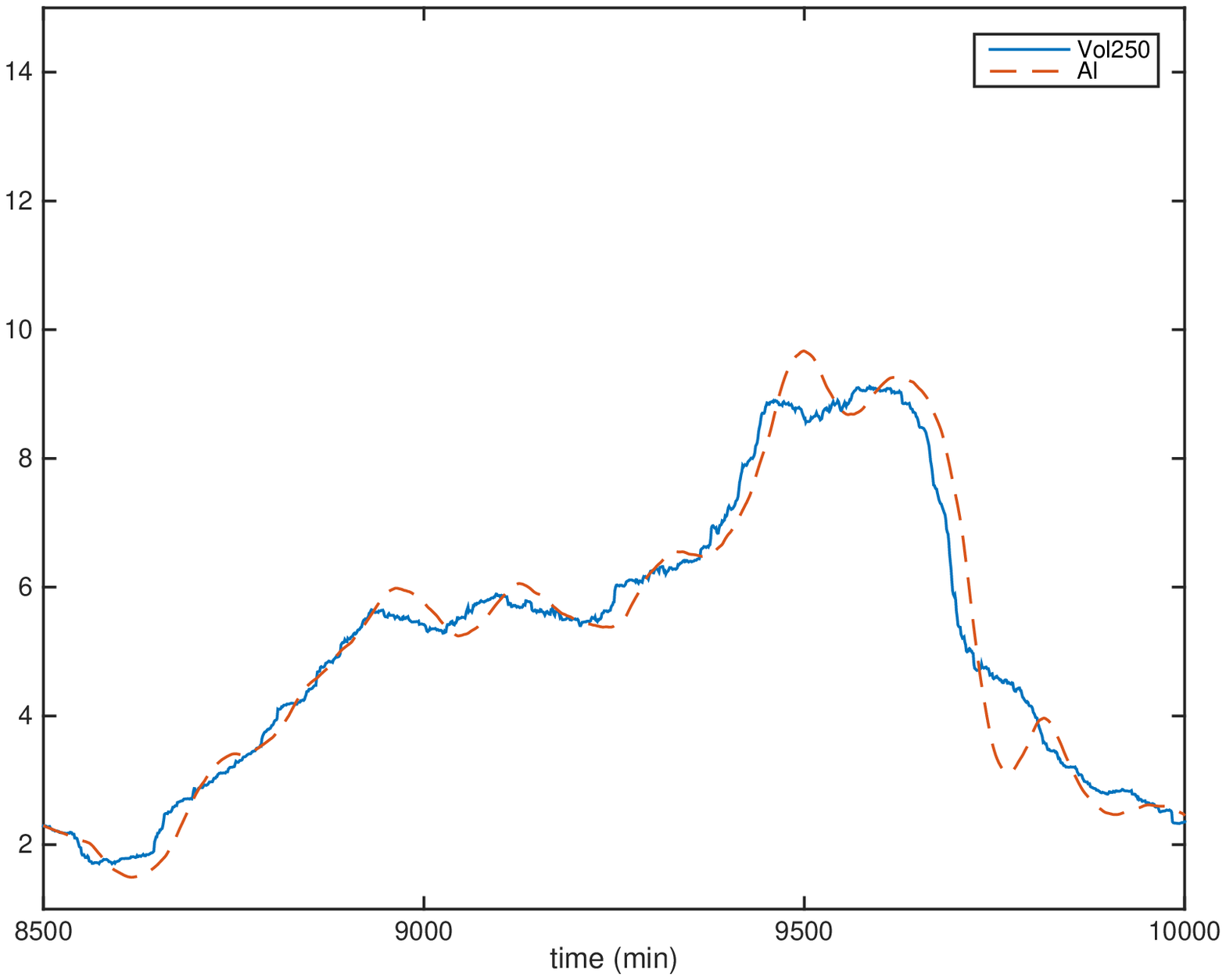}}}%
\subfigure[Zoom 2]{
\resizebox*{5.915cm}{!}{\includegraphics{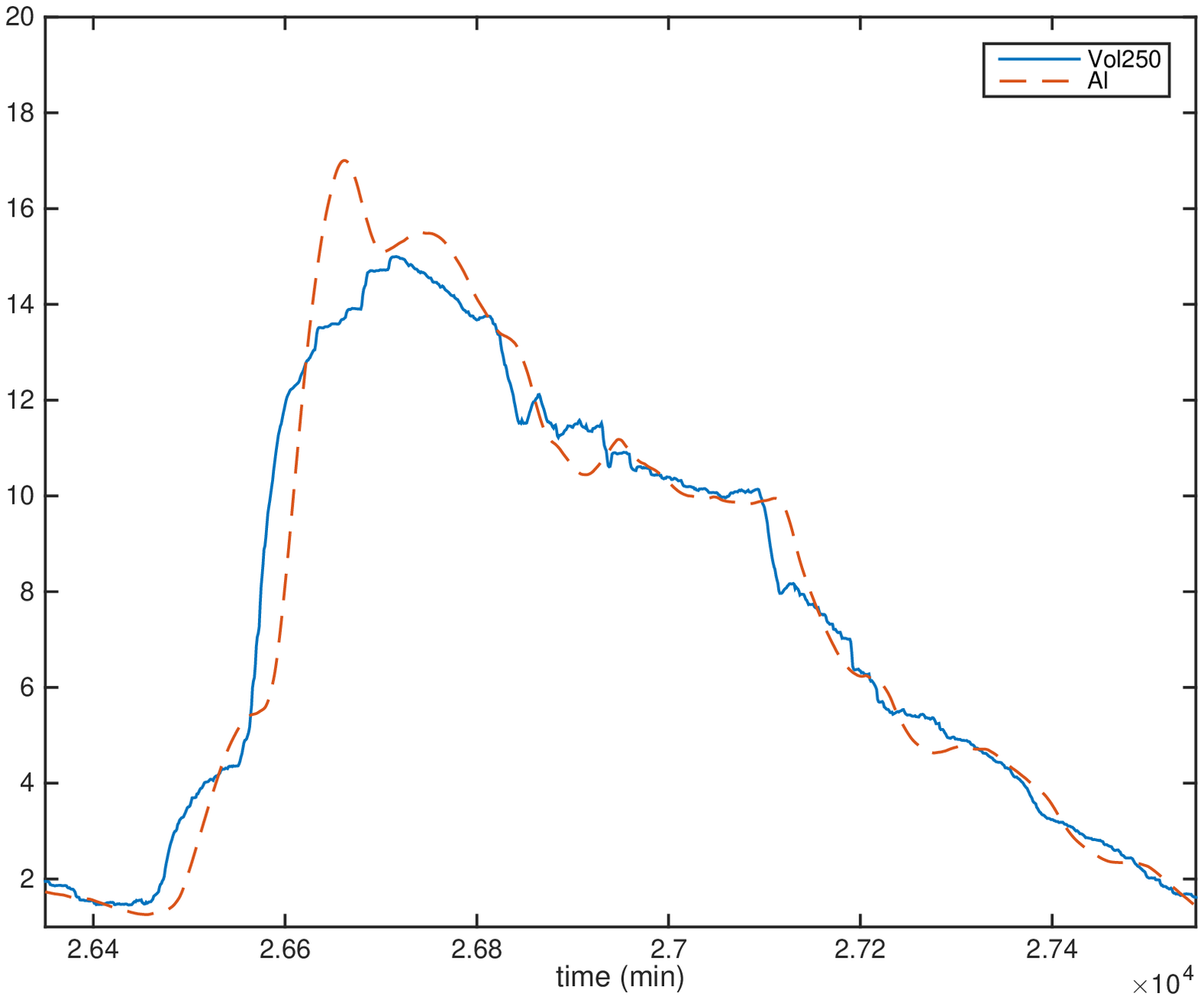}}}%
\caption{Forcasting volatility}%
\label{VP}
\end{center}
\end{figure*}

Figure \ref{Ts} displays the various trends, which are computed via the non-causal mean \eqref{mean}, for three time scales: $100$, $250$, or $500$ minutes. Note that larger is 
the time scale smoother is the trend. On the other hand Figure \ref{Vs} shows a volatility increase. 
Due to a lack of space, only forecasting volatility via Formula \eqref{A1}  for a $15$ minutes time horizon is displayed in Figure \ref{VP}, where the middle value, \textit{i.e.}, $250$ minutes, is  utilized for calculating the trend.  The results are rather good.


\section{Conclusion}\label{conclusion}
Although encouraging our preliminary results need not only to be further developed but also to be compared 
with other existing approaches. Let us emphasize that such comparisons began to be discussed for short-term meteorological forecasts by \cite{paris} and \cite{voyant}. Our methods were easier to implement 
and much less demanding in terms of historical data. 
For a deeper study of reliability and risk, see \cite{troyes} where the notion of \emph{confidence bands} may be extended to traffic management in a straightforward way.

\begin{ack}
The \textit{Cerema} (\textit{\underline{C}entre d'\'etudes et d'\underline{e}xpertise 
sur les \underline{r}isques, l'\underline{e}nvironnement, la \underline{m}obilit\'e et l'\underline{a}ménage\-ment}) provided the authors with the necessary data for the highway A25. This highway is managed by the DIRN (\textit{\underline{D}irection \underline{I}nterd\'epartementale des \underline{R}outes \underline{N}ord}) via the \textit{ALLEGRO} (\textit{\underline{A}gglom\'eration li\underline{LL}oise 
\underline{E}xploitation \underline{G}estion de la \underline{RO}ute}) system. 
\end{ack}

\bibliography{ifacconf}             

\end{document}